\documentclass[11pt]{revtex4}
\usepackage{booktabs}
\usepackage{titlesec}
\usepackage{graphicx}
\usepackage{setspace}

\begin{document}

\title{Improving Students¡¯ Understanding of Quantum Measurement \\ Part 1: Investigation of Difficulties}

\author {Guangtian Zhu}
\affiliation {School of Education Science, East China Normal University, Shanghai, China, 200062}

\author {Chandralekha Singh}
\affiliation {Department of Physics \& Astronomy, University of Pittsburgh, Pittsburgh, PA, 15260}

\begin{abstract}
We describe the difficulties that advanced undergraduate and graduate students have with quantum measurement within the standard interpretation of quantum mechanics. We explore the possible origins of these difficulties by analyzing student responses to questions from both surveys and interviews. Results from this research are applied to develop research-based learning tutorials to improve students' understanding of quantum measurement.
\end{abstract}

\maketitle


\section{INTRODUCTION}

Quantum mechanics is a particularly challenging subject, even for the advanced students [1-4]. These difficulties have been described in a number of investigations [5-12]. Based on these findings, we are developing a set of research-based learning tools to reduce students' difficulties and help them develop a solid grasp of quantum mechanics [13-15]. This paper (Part 1) is the first of two in which we discuss the investigation of students' difficulties with quantum measurement. This investigation was conducted with the undergraduate and graduate students at the University of Pittsburgh (Pitt) and other universities by administering written tests and by conducting in-depth individual interviews with a subset of students. The development of the research-based learning tools and the preliminary evaluation of students' performance after using the learning tools is described in the second of the two papers (Part 2) [16].

The standard formalism of quantum measurement (which is taught to the undergraduate and graduate students universally) is quite different from classical mechanics, where position and momentum of a particle evolve in a deterministic manner based upon the interactions [1]. In quantum mechanics, position, momentum and other observables are in general not well-defined for a given state of a quantum system. The time-dependent Schroeinger equation (TDSE) governs the time evolution of the state which can be written as a linear superposition of a complete set of eigenstates of any hermitian operator corresponding to a physical observable. The state of the system evolves in a deterministic manner depending on the Hamiltonian of the system. According to the Copenhagen interpretation, quantum measurement would instantaneously collapse the wavefunction (or the state of the system) to an eigenstate of the operator corresponding to the physical observable measured and the measured value is the corresponding eigenvalue. For example, in an ideal measurement, if we measure the position of a quantum particle in a one-dimensional (1D) infinite square well, its wavefunction will collapse to a position eigenfunction which is a delta function in the position representation. If we measure its energy instead, the wavefunction of the system will collapse into an energy eigenfunction, which is a sinusoidal function inside the 1D well and goes to zero at the two boundaries (and is zero everywhere outside the well).

The eigenvalue spectrum of an operator can either be discrete or continuous or a combination of the two. In an \emph{N} dimensional Hilbert space, an operator $ \hat Q\ $ corresponding to a physical observable $Q$ with discrete eigenvalue spectrum has N eigenvalues  $\ q_n $ and corresponding eigenstates $\ \left | {q_n } \right\rangle\ $. The state of the system at a given time, $\ \left | {\Psi (t)} \right\rangle\ $, can be written as a linear superposition of a complete set of eigenstates of $\ \left | {q_n } \right\rangle\ $. By projecting the wavefunction of the system $\ \left| {\Psi (t)} \right\rangle\ $ at time $t$ onto an eigenstate  $\ \left | {q_n } \right\rangle\ $ of the operator $\ \hat Q\ $, we can find the probability $\ \left| {\left\langle {{q_n }}\mathrel{\left | {\vphantom {{q_n }{\Psi (t)}}}\right. \kern-\nulldelimiterspace}{{\Psi (t)}}\right\rangle }\right|^2\ $ of obtaining $\ q_n$ when the observable $Q$ is measured at time $t$.

After the measurement of the observable $Q$, the time-evolution of the state of the system, which is an eigenstate of $\ \hat Q\ $ right after the measurement, is again governed by the TDSE. Right after the measurement of energy, the state of the system collapses into the same energy eigenstate, and the probability density does not change with time (we only focus on time-independent Hamiltonians) since the only change in the wavefunction with time is an overall time-dependent phase factor. If the system is initially in an energy eigenstate at time $t=0$ and we measure an arbitrary physical observable $Q$ after a time $t$, the probability of obtaining an eigenvalue   will be time-independent since the system was still in an energy eigenstate at time $t$ at the instant the measurement of $Q$ was performed. Therefore, the energy eigenstates are called the stationary states. On the other hand, a measurement of position would collapse the system into a position eigenstate at the instant the measurement is made. However, a position eigenstate is a linear superposition of the energy eigenstates and the different energy eigenstates in the linear superposition will evolve with different time-dependent phase factors. Therefore, the probability density after position measurement will change with time. In this case, the probability of measuring a particular value of energy will be time-independent but the probability of measuring another physical observable whose operator does not commute with the Hamiltonian will depend on time.

\section{INVESTIGATION OF STUDENTS' DIFFICULTIES}

The goal of the investigation was to examine students' difficulties with quantum measurement after traditional instruction. The topics included in the investigation, such as the measurement outcomes, probability of obtaining an eigenvalue, stationary states and eigenstates, etc., were all covered in the traditional instruction of quantum mechanics. The investigation was carried out over several years. For example, students were given the questions as part of the concept tests, quizzes or tests depending upon the instructor's preference. Therefore, the number of students who answered a particular question varies. To simplify the mathematics and focus on the concepts related to quantum measurement, we often used the model of a 1D infinite square well during the investigation. Both open-ended questions and multiple-choice questions were administered to probe students' difficulties. We also had informal discussions with a subset of students who took the written test and formally interviewed some students to get a better understanding of students' reasoning process.
\\

\textbf{ A. Difficulty in Distinguishing between Eigenstates of Operators corresponding to Different Observables}
\\
The measurement of a physical observable collapses the wavefunction of the quantum system into an eigenstate of the corresponding operator. Many students have difficulties distinguishing between energy eigenstates and the eigenstates of other physical observables. To investigate the pervasiveness of this difficulty in distinguishing between the eigenstates of different physical observables, one of the multiple choice questions administered to the students was the following:
\begin{itemize}
\item \emph{Choose all of the following statements that are correct:}
\end{itemize}
\emph{(1) The stationary states refer to the eigenstates of any operator corresponding to a physical observable.\\
(2) If a system is in an eigenstate of any operator that corresponds to a physical observable, it stays in that state unless an external perturbation is applied.\\
(3) If a system is in an energy eigenstate at time $t=0$, it stays in the energy eigenstate unless an external perturbation is applied.\\
$A$. 1 only \ \ \  $B$. 3 only \ \ \ $C$. 1 and 3 only \ \ \
$D$. 2 and 3 only \ \ \ $E$. all of the above}
\\
\\
The correct answer is $B$ (3 only). In statement (1), the stationary states should refer to the energy eigenstates only. A complete set of eigenstates of an arbitrary operator $\ \hat Q\ $  cannot be stationary states if $\ \hat Q\ $ does not commute with the Hamiltonian operator $\ \hat H\ $. However, out of 10 students randomly selected from a junior-senior level quantum mechanics class, none of them gave the correct answer after traditional instruction. The distribution of students' answers is shown in Table 1. The most common incorrect choice was $E$ (all of the above). Nearly half of the students thought that all three statements were correct because they had difficulty in differentiating between the related concepts of stationary states and eigenstates of other observables. Some students selected choice $A$ (1 only) which is interesting because one may expect that students who claimed statement (1) was correct and understood why a stationary state is called so may think that statement (2) is correct as well. In particular, for students who claimed statement (1) is correct, statement (2) may be considered ``a system in a stationary state stays in that state unless an external perturbation is applied", which described the property of stationary state. However, students who selected choice $A$ did not relate the stationary state with the special nature of the time evolution in that state.
\\

\begin{table}
  \centering
 \caption[1]{\parbox[t]{9.5cm}{The choice distribution of 10 students answering the question about stationary state and eigenstate after traditional instruction.}}
  \begin{tabular}{|c|c|c|c|c|c|c|}
    \hline
 Answers    & \ \ \ A \ \ \ & B(correct)  & \ \ \ C \ \ \ & \ \ \ D \ \ \ & \ \ \ E \ \ \ & No Answer \\
    \hline
Percentage  & \ \ \ 20$\%$ \ \ \ & 0$\%$	& \ \ \ 0$\%$ \ \ \ & \ \ \ 10$\%$ \ \ \ & \ \ \ 50$\%$	\ \ \ & 20$\%$ \\
    \hline
  \end{tabular}
\end{table}

\textbf{ B. Difficulty with possible outcomes of a measurement and the expectation value of the measurement result}
\\
The following is an example of a multiple choice question which was administered to investigate students' understanding of the possible outcomes of a measurement for a given state of a particle in a 1D infinite square well when the measurement is performed.$\ \psi _1 (x)\ $ and  $\ \psi _2 (x)\ $ are the ground state and first excited state wavefunctions. \\
\begin{itemize}
\item \emph{An electron is in the state given by $\ \frac{{\psi _1 (x) + \psi _2 (x)}}{{\sqrt 2 }}\ $. Which one of the following outcomes could you obtain if you measure the energy of the electron?}
\end{itemize}
\emph{A. $ \ E_1  + E_2\ $ \\
B. $ \ (E_1  + E_2 )/2\ $\\
C. Either  $\ E_1\ $ or $\ E_2\ $ \\
D. Any of $\ E_n\ $ (n=1,2,3,¡­) \\
E. Any value between $\ E_1\ $ and $\ E_2\ $}\\

Because the energy eigenstates $\ \left| {\psi _n } \right\rangle\ $ are orthogonal to each other, $\ \left| {\left\langle {{\psi _n }} \mathrel{\left | {\vphantom {{\psi _n } \Psi }} \right. \kern-\nulldelimiterspace} {\Psi } \right\rangle } \right|^2  = 1/2\ $ for both $n=1$ and $n=2$ and $\ \left| {\left\langle {{\psi _n }} \mathrel{\left | {\vphantom {{\psi _n } \Psi }} \right. \kern-\nulldelimiterspace} {\Psi } \right\rangle } \right|^2  = 0\ $ for all the other energy eigenstates $E_n$ ($n>2$). Therefore, we can only obtain $\ E_1\ $ or $\ E_2\ $ with equal probability but no other energy. The distribution of students' answers is shown in Table 2. Six out of fifteen students in a junior-senior quantum mechanics class chose the correct answer $C$ (either $\ E_1\ $ or $\ E_2\ $). The most common incorrect choice selected by 27$\%$ of the 15 students was $B$ ($\ (E_1 + E_2)/2\ $) which actually represents the expectation value of energy. Students mistakenly claimed that the expectation value is the measured value of energy. Informal discussions with individual students and individual think-aloud interviews [17] indicated that many students were not only confused about the distinction between individual measurements and expectation values, they also had difficulty distinguishing between the probability of measuring a particular value of an observable in a given state and the measured value or the expectation value. For example, during individual interviews, students often wrote $\ \left\langle {\psi _n } \right|\hat H\left| {\psi _n } \right\rangle\ $ or even $\ \left\langle \Psi  \right|\hat H\left| \Psi  \right\rangle\ $ as the probability of measuring  $\ E_n\ $ in the state $\ \left| \Psi  \right\rangle\ $. When these students were explicitly asked to compare their expressions for the probability of measuring a particular value of energy and the expectation value of energy, some students appeared concerned. They recognized that these two concepts were different but they still struggled to distinguish these concepts. They could not write an expression for the probability of measuring $\ E_n\ $ either using the Dirac notation or in the position space representation using the integral form.

Discussions with students and individual interviews suggest that some of them had difficulty in differentiating between the probability of measuring each possible value of an observable and the expectation value of that observable in a given state. Since the expectation value in a given state equals the average of a large number of measurements of that observable on identically prepared systems, it is equal to the sum of the eigenvalues of the corresponding operator times their probabilities in the given state. Many students had difficulty with the statistical interpretation of the expectation value of $Q$ as the average of a large number of measurements on identically prepared systems in state $|\Psi\rangle$. For example, a survey question which was administered to 202 graduate students from seven universities illustrates it as shown below [10]:
\begin{itemize}
\item \emph{The wavefunction of an electron in a 1D infinite square well of width $ \ a \ $ at time t=0 is given by  $ \ \Psi (x,t = 0) = \sqrt {2/7} \psi _1 (x) + \sqrt {5/7} \psi _2 (x)\ $. Answer the following questions.}
\end{itemize}
\emph{(a) You measure the energy of an electron at time t=0. Write down the possible values of the energy and the probability of measuring each.\\
(b) Calculate the expectation value of the energy in the state $ \ \Psi (x,t)\ $.}

 67$\%$ of the graduate students answered question (a) correctly and 7$\%$ of them were confused about the distinction between the energy eigenvalues and the expectation value of energy. However, only 39$\%$ of the students provided the correct response for question (b) above. Many students who could calculate the probability for measuring each energy in question (b) did not use the probabilities to find the expectation value. Some of them tried to find the expectation value by sandwiching the Hamiltonian with the state of the system (i.e., $ \left\langle \Psi \right|\hat H\left| \Psi  \right\rangle\ $) which is correct and some even wrote down correct corresponding integrals but then struggled with the calculation.\\

 \begin{table}
  \centering
 \caption[1]{\parbox[t]{9.5cm}{The choice distribution of 15 students answering the question about energy measurement outcome after traditional instruction.}}
  \begin{tabular}{|c|c|c|c|c|c|c|}
    \hline
 Answers    & \ \ \ A \ \ \ & \ \ \ B \ \ \  & C(correct) & \ \ \ D \ \ \  & \ \ \ E  \ \ \ & No Answer \\
    \hline
Percentage  & \ \ \ 13$\%$ \ \ \	& \ \ \ 27$\%$ \ \ \ & 40$\%$ & \ \ \ 7$\%$ \ \ \	& \ \ \ 7$\%$ \ \ \ & 6$\%$ \\
    \hline
  \end{tabular}
\end{table}

\textbf{C. Difficulty with the probability of measuring energy}

When we explicitly asked students to find the probability of obtaining energy $\ E_2\ $ for the state $\ \frac{{\left| {\psi _1 } \right\rangle  + \left| {\psi _2 } \right\rangle }}{{\sqrt 2 }}\ $ in a 1D infinite square well, many of them could provide the correct answer 1/2 by observing the coefficients. To evaluate whether students could calculate the probability of measuring a particular value of energy by projecting the state vector along the corresponding energy eigenstate for the case where the wave function is not written explicitly in terms of a linear superposition of energy eigenstates, the following question about a triangle shaped wavefunction in a 1D infinite square well was administered:
\begin{itemize}
\item \emph{The state of an electron at t=0 is given by $\ \Psi (x) = Ax\ $ when $\ 0 < x < \frac{a}{2}\ $,$\ \Psi (x) = A(a - x)\ $ when $\ \frac{a}{2} \le x < a\ $ and  $\ \Psi (x) = 0\ $  elsewhere. Here $A$ is the normalization constant. What is the probability that an energy measurement at time t=0 yields energy $E_2$? (If there is an integral in your expression for the probability, you need not evaluate the integral but set it up properly with appropriate limits. Ignore the fact that the first derivative of the wavefunction is not continuous.)}
\end{itemize}

Unlike the state $\ \left( {\left| {\psi _1 } \right\rangle  + \left| {\psi _2 } \right\rangle } \right)/\sqrt 2\ $ which is composed of only two energy eigenstates, the triangle function state $\ \left| \Psi  \right\rangle\ $ (or $\ \Psi (x)\ $ in the position space) is a superposition of infinitely many energy eigenstates, i.e., $\ \left| \Psi  \right\rangle  = \sum\limits_{n = 1}^\infty  {c_n \left| {\psi _n } \right\rangle }\ $. The expansion coefficient equals $\langle \psi_n|\Psi\rangle = \int\limits^{+\infty}_{-\infty} \psi ^\ast_n (x)\Psi(x)dx$ and $|c_n|^2$ is the probability of obtaining energy $E_n$ when energy is measured for the state $\ \left| \Psi \right\rangle\ $. Thus, to answer this question correctly, students need to write $\ \left| \Psi  \right\rangle\ $ as a linear superposition of $\ \left| {\psi _n } \right\rangle\ $  and find the component of $\ \left| \Psi\right\rangle\ $ along $\ \left| {\psi _n } \right\rangle\ $.

Only one student out of fifteen provided the correct answer and some students left this question blank. Other students had two common mistakes. Twenty percent of the students wrote down the energy expectation value $\ \left\langle \Psi  \right|\hat H\left| \Psi  \right\rangle\ $ to represent the energy measurement probability. In further informal discussions and formal interviews with some students, we asked how the expression $\ \left\langle \Psi  \right|\hat H\left| \Psi  \right\rangle\ $ which only involved state $\ \left| \Psi  \right\rangle\ $ would favor energy $\ E_2\ $ over any other energy. Some of the students then changed their answers to  $\ \left\langle {\psi _2 } \right|\hat H\left| \Psi  \right\rangle\ $ which was still incorrect. Another 27$\%$ of the students claimed that the ``probability" of measuring any physical observable was represented by $\ \left| {\Psi (x)} \right|^2\ $ according to the interpretation of wavefunction. These students were confusing the probability density for measuring position with the probability of measuring other physical observables such as energy.

A similar multiple-choice question about a parabola shaped wavefunction was administered to 76 students in six universities as shown below:
\begin{itemize}
\item \emph{Consider the following wavefunction for a 1D infinite square well: $\ \Psi (x) = Ax(a - x)\ $ for $\ 0 \le x \le a\ $ and $\ \Psi (x) = 0\ $ otherwise. $A$ is a normalization constant. Which one of the following expressions correctly represents the probability of measuring the energy $\ E_n\ $ for the state $\ \Psi (x)\ $?}
\end{itemize}
A. $\ \left| {\int\limits_0^a {\psi _n^* (x)\hat H\Psi (x)dx} } \right|^2\ $ \\
B. $\ \left| {\int\limits_0^a {\psi _n^* (x)\Psi (x)dx} } \right|^2\ $ \\
C. $\ \left| {\psi _n^* (x)\hat H\Psi (x)} \right|^2\ $ \\
D. $\ \left| {\psi _n^* (x)\Psi (x)} \right|^2\ $ \\
E. $\ \left| {\Psi (x)} \right|^2\ $ \\
Among the 76 students, 61 were junior/senior undergraduate students and the others were first year graduate students in physics department. The distributions of the undergraduate and graduate students' answers are listed in Table 3. About one third of both the graduate and undergraduate students chose the correct answer $B$. The undergraduate students tended to include the Hamiltonian operator in calculating the probability of measuring a particular energy eigenvalue. For example, 49$\%$ of the undergraduate students incorrectly selected the distractor option A which is an equivalent expression for $\ \left| {\left\langle {\psi _2 } \right|\hat H\left| \Psi  \right\rangle } \right|^2$. However, the first year graduate students were more likely to neglect the integral part in calculating the measurement probability. 5 out of 15 graduate students mistakenly chose the option $D$, $\ \left| {\psi _n^* (x)\Psi (x)} \right|^2$, but only 2$\%$ of the undergraduate students made such mistake.

Another multiple choice question given to the same 76 students asked about the energy measurement outcomes for the state $\ \sqrt {4/7} \left| {\psi _1 } \right\rangle  + \sqrt {3/7} \left| {\psi _2 } \right\rangle$. 55$\%$ of all the 76 students provided the correct answer. 21$\%$ of the students incorrectly claimed that other energies $\ E_n\ $ besides $\ E_1\ $ and $\ E_2\ $ could also be obtained but the probability of measuring $\ E_1\ $ would be largest. Another 12$\%$ of the students thought that all the possible energies $\ E_n\ $ can be measured with the same probability.\\

\begin{table}
  \centering
 \caption[1]{\parbox[t]{9.5cm}{The choice distributions of 61 undergraduate students and 15 graduate students answering the question about energy measurement probability.}}

  \begin{tabular}{|c|c|c|c|c|c|c|}
    \hline
 Answers    & \ \ \ A \ \ \ & B(correct)  & \ \ \ C \ \ \ & \ \ \ D \ \ \ & \ \ \ E \ \ \ & No Answer \\
    \hline
Undergrad.  & \ \ \ 49$\%$ \ \ \ & 31$\%$	& \ \ \ 10$\%$ \ \ \ & \ \ \ 2$\%$ \ \ \ & \ \ \ 5$\%$	\ \ \ & 3$\%$ \\
    \hline
Graduate  & \ \ \ 27$\%$ \ \ \ & 27$\%$	& \ \ \ 0$\%$ \ \ \ & \ \ \ 33$\%$ \ \ \ & \ \ \ 13$\%$	\ \ \ & 0$\%$ \\
    \hline
  \end{tabular}
\end{table}

\textbf{D. Difficulties with the time development of the wavefunction after the measurement of an observable}
\\
Within the Copenhagen interpretation of quantum mechanics, the measurement of an observable is treated separately from the ``normal" time-evolution of the system according to TDSE. When a measurement is performed, the state of the system instantaneously collapses to an eigenstate of the operator corresponding to the observable measured after which the system will evolve normally according to the TDSE. We investigated students' understanding of the time-development of the wavefunction according to the TDSE after the measurement of an observable by asking 15 students the following question about consecutive position measurement for a 1D infinite square well:
\begin{itemize}
\item \emph{If you make a measurement of position on an electron in the ground state of a 1D infinite square well and wait for a long time before making a second measurement of position, do you expect the outcome to be the same in the two measurements? Explain.}
\end{itemize}

To correctly answer this question, students must know the following: (1) the ground state wavefunction will collapse into a position eigenfunction (a delta function in position) after the first position measurement; (2) the position eigenfunction is not a stationary state wavefunction so the wavefunction will evolve in time in a non-trivial manner and it will not in general be found in a position eigenstate after a time $t$. Therefore, after a long time, the second measurement of position in general will yield a different value from the first measurement. We note however that in an infinite square well, the time evolution of the system is such that the wave function repeats itself with a certain periodicity.\\

\textbf{Difficulity D.1: System remains in the energy eigenstate after a position measurement}

In response to this question, some students thought that the system will be in the ground state after both the first and the second position measurements. Informal discussions with some students and formal interviews with a handful of students suggest that those with these types of responses often did not realize the difference between an energy eigenstate and a position eigenstate. They claimed that if the system is in the ground state, it will remain in that state. Students who were explicitly asked what would happen if the initial state before the measurement was the first excited state (which is also an energy eigenstate) typically responded that it will remain in that state since even that state is an ``eigenstate". In the written survey, only one out of fifteen students explicitly mentioned the wavefunction collapse after the first position measurement. However, his response was ``¡­the wavefunction collapses into the measured state" and he did not elaborate that the ``measured state" is actually a position eigenstate.\\

\textbf{Difficulty D.2: System stays in the position eigenstate at any time after a position measurement}

Some students claimed that after the first position measurement the system gets ``stuck" in a position eigenstate and did not know that the position eigenfunction (unlike the energy eigenfunctions for a time-independent Hamiltonian) evolves in time in a non-trivial manner and the system does not remain a position eigenfunction for all future time $t$. These students claimed that the second position measurement will yield the same value as the first one unless there was an ``outside disturbance". Only two out of fifteen students mentioned the correct time evolution of the quantum mechanical system after the position measurement.\\

\textbf{Difficulty D.3: System finally goes back to the initial state}

Students were also asked another series of questions about measurement when the initial state of the system at time $t=0$ is $\ \Psi (x,0) = \sqrt {2/7} \psi _1 (x) + \sqrt {5/7} \psi _2 (x)\ $ for an electron confined in a 1D infinite square well as follows:

\begin{itemize}
\item \emph{Q1. If the energy measurement yields $\ 4\pi ^2 \hbar ^2 /(2ma^2 )$, what is the wavefunction right after the measurement?}
\end{itemize}

\begin{itemize}
\item \emph{Q2. Immediately after the energy measurement in Q1, you measure the position of the electron. What possible values could you obtain and what is the probability of each?}
\end{itemize}

\begin{itemize}
\item \emph{Q3. After the position measurement in Q2, you wait for time $t>0$ and measure the position again. Would the probability of measuring each possible value different from Q2?}
\end{itemize}

Q1 which asks about the state of the system long after the energy measurement (instead of immediately after the measurement as in the open-end question) has been given as a multiple-choice question to 76 students from 6 universities. An analysis of the student responses suggests that 20$\%$ of the students did not know that the wavefunction would collapse at the instant the energy was measured. Also, 36$\%$ of the students thought the wavefunction will collapse upon energy measurement but finally evolved back to the initial state $\ \sqrt {2/7} \psi _1 (x) + \sqrt {5/7} \psi _2 (x)\ $ long time after the measurement. During the individual interview, a student said, ``¡­it's like tossing a coin. You can get either head or tail after the measurement. But when you make another measurement, it goes back to a coin (with two sides)." Such a statement also indicates that the student made an inappropriate transfer of a classical probability concept to quantum probability.\\

\textbf{Difficulty D.4: Probability density for position measurement}

Born's probabilistic interpretation of the wavefunction can also be confusing for students. In Q2 above, the wavefunction of the system before the position measurement is the energy eigenstate $\psi _2 (x) = \sqrt {2/a} \sin (2\pi x/a)$. We expected students to note that one can measure position values between $x=0$ and $x=a$ (except $x=0$, $a/2$, and $a$ where the wavefunction is zero), and according to Born's interpretation,$\ \left| {\psi _2 } \right|^2 dx\ $ gives the probability of finding the particle in a narrow range between $x$ and $x+dx$. However, only 38$\%$ of students provided the correct response. Partial responses were considered correct for tallying purposes if students wrote anything that was correct related to the above wavefunction, e.g., ``The probability of finding the electron is highest at $a/4$ and $3a/4$", ``The probability of finding the electron is non-zero only in the well", etc.[18]\\

Eleven percent of the students tried to find the expectation value of position instead of the probability of finding the electron at a given position. They wrote the expectation value of position in terms of an integral involving the wavefunction. Many of them explicitly wrote that $\ probability = (2/a)\int_0^a x \sin ^2 (2\pi x/a)dx\ $ and claimed that instead of the expectation value they were calculating the probability of measuring the position of the electron.

During the interview [18], one student said (and wrote on paper) that the probability of position measurement is $\int {x\left| \psi\right|^2 dx}\ $ ($\ \psi  = \psi _2\ $ in Q2 above). When the interviewer asked why $ \ \left| \psi  \right|^2\ $ should be multiplied with $x$ and if there is any significance of $\ \left| \psi  \right|^2 dx\ $ alone without multiplying it by $x$, the student said, ``$\left| \psi  \right|^2\ $ gives the probability of the wavefunction being at a given position and if you multiply it by $x$ you get the probability of measuring (student's emphasis) the position $x$". When the student was asked questions about the meaning of the ``wavefunction being at a given position", and the purpose of the integral and its limits, the student was unsure. He said that the reason he wrote the integral is because $\ x\left| \psi  \right|^2 dx\ $ without an integral looked strange to him. Similar confusion about probability in classical physics situations have been found [19].\\
\textbf{Difficulty D.5: Use of classical language to describe time evolution of quantum systems}

Out of the ten students who were given Q3 above, none of them could answer it correctly though it assesses the same concepts as in the \emph{consecutive position measurement question} discussed earlier. In the \emph{consecutive position measurement question}, some students used a classical description to answer the question about the time-evolution after the measurement such as ``the electron moves around". Discussions with individual students and interviews suggest that such classical responses reflect students' discomfort describing the time evolution of a quantum system in terms of the time-development of wavefunction.\\

\textbf{E. Incorrectly believing that an operator acting on a state corresponds to a measurement of the corresponding observable}

One of the questions on a survey given to more than 200 graduate students asked them to consider the following statement[10]: ``By definition, the Hamiltonian acting on any allowed (possible) state of the system $\ \left| \psi  \right\rangle\ $ will give the same state back, i.e. $\ \hat H\left| \psi \right\rangle  = E\left| \psi  \right\rangle\ $,where $E$ is the energy of the system." Students were asked to explain why they agree or disagree with this statement. We expected students to disagree with the statement and note that it is only true if $\ \left| \psi  \right\rangle\ $ is a stationary state. In general,$\ \left| \psi  \right\rangle  = \sum\limits_{n = 1}^\infty  {C_n \left| {\psi _n } \right\rangle }\ $ where $\ \left| {\psi _n } \right\rangle\ $ are the stationary states and $\ C_n  = \left\langle {{\psi _n }} \mathrel{\left | {\vphantom {{\psi _n } \psi }} \right. \kern-\nulldelimiterspace} {\psi } \right\rangle\ $. Then, $\ \hat H\left| \psi  \right\rangle= \sum\limits_{n=1}^\infty  {C_n E_n \left| {\psi _n } \right\rangle } \ne E\left| \psi\right\rangle\ $.

Eleven percent of the students answering this question incorrectly claimed that any statement involving a Hamiltonian operator acting on a state is a statement about the measurement of energy. Some of these students who incorrectly claimed that $\ \hat H\left| \psi \right\rangle  = E\left| \psi  \right\rangle\ $ is a statement about energy measurement agreed with the statement while others disagreed. Those who disagreed often claimed that $\ \hat H\left| \psi  \right\rangle  = E_n \left| {\psi _n } \right\rangle\ $ because as soon as $\ \hat H\ $ acts on $|\psi\rangle$, the wavefunction will collapse into one of the stationary states $\ \left| {\psi _n } \right\rangle\ $ and the corresponding energy $\ E_n\ $ will be measured. The following are two typical responses in this category:
\begin{itemize}
\item Disagree. Hamiltonian acting on a state (measurement of energy) will return an energy eigenstate.
\end{itemize}
\begin{itemize}
\item When $\ \left| \psi  \right\rangle\ $ is a superposition state and $\ \hat H\ $ acts on $\ \left| \psi  \right\rangle\ $, $\ \left| \psi  \right\rangle\ $  evolutes to one of the $\ \left| {\psi _n } \right\rangle\ $ so we have $\ \hat H\left| \psi  \right\rangle  = E_n \left| {\psi _n } \right\rangle\ $.
\end{itemize}

Formal interviews, informal discussions and written reasonings suggest that these students often believed that the measurement of \emph{any} physical observable in a particular state is achieved by acting with the corresponding operator on the state. The incorrect notions expressed above are often over-generalizations of the fact that \emph{after} the measurement of energy, the system is in a stationary state so $\ \hat H\left| {\psi _n } \right\rangle  = E_n \left| {\psi _n } \right\rangle\ $ and students felt that there should be an equation describing the collapse of the wave function.

Individual interviews related to this question suggest that some students believed that whenever an operator $\ \hat Q\ $ corresponding to a physical observable $Q$ acts on any state $|\psi\rangle$, it will either yield a corresponding eigenvalue $\ \lambda\ $ and the same state back, i.e., $\ \hat Q\left| \psi  \right\rangle  = \lambda \left| \psi  \right\rangle\ $ or yield $\ \hat Q\left| \psi  \right\rangle  = \lambda _n \left| {\phi _n } \right\rangle\ $ where $\ \left| {\phi _n } \right\rangle\ $ is the $n^{th}$ eigenstate of $\ \hat Q\ $ in which the system collapses and $\ \lambda _n\ $ is the corresponding eigenvalue (but actually, $\ \hat Q\left| {\phi _n } \right\rangle  = \lambda _n \left| {\phi _n } \right\rangle\ $).

We further explored this issue by asking 17 and 15 graduate students at the end of their first semester and second semester graduate level quantum mechanics course the following question. 15 graduate students were the same in both semesters.
\begin{itemize}
\item \emph{Consider the following conversation between Andy and Caroline about the measurement of an observable $\ Q\ $ for a system in a state $\ \left| \psi  \right\rangle\ $ which is not an eigenstate of $\ \hat Q\ $:}
\end{itemize}
\emph{\textbf{Andy}: When an operator $\ \hat Q\ $  corresponding to a physical observable $\ Q\ $  acts on the state $\ \left| \psi  \right\rangle\ $ it corresponds to a measurement of that observable. Therefore, $\ \hat Q\left| \psi  \right\rangle  = q\left| \psi  \right\rangle\ $ where q is the observed value.\\
\textbf{Caroline}: No. The measurement collapses the state so $\ \hat Q\left| \psi \right\rangle = q\left| {\psi _q } \right\rangle\ $ where $\ \left| {\psi _q }\right\rangle\ $ on the right hand side of the equation is an eigenstate of $\ \hat Q\ $ with eigenvalue $q$.\\
 With whom do you agree?\\}
\emph{A.	Agree with Caroline only\\
B.	Agree with Andy only\\
C.	Agree with neither\\
D.	Agree with both\\
E.	The answer depends on the observable $Q$.}
\\
We note that the question was not posed as a multiple-choice question at the end of the first semester course but students were asked to explain whom if any they agreed with and why. There was a brief discussion of the correct response to the question after administering the survey in which this question was asked. At the end of the first semester course, 12$\%$ of the students agreed with Andy, 47$\%$ with Caroline, 29$\%$ with neither (correct response) and 12$\%$ provided no response. In the second semester, the concepts about measurement were not explicitly emphasized in the course of Quantum Mechanics II. At the end of the second semester course, the same question in the multiple-choice form was administered. This time, 20$\%$ of the students chose the distractor $A$, which is lower than the 47$\%$ in the first semester. However, about the same percentage of students in both semesters thought that the operator corresponding to an observable acting on any quantum state gives the eigenvalue, i.e.,$\ \hat Q\left| \psi  \right\rangle  = q\left| \psi  \right\rangle\ $ . The comparison of students' answer distribution is listed in Table 4. 13$\%$ of the students agreed with Andy, 20$\%$ with Caroline, 7$\%$ with both and 53$\%$ with neither (correct response). While the percentage of correct response increased significantly from the first to the second administration, many students still had difficulty with this concept. Earlier, the version of this question not in the multiple-choice format was posed to 37 graduate students at the beginning of their graduate level quantum mechanics course (not the same students as those who answered it at the end of the first and second semester of their graduate level quantum mechanics course). In that group, 24$\%$ of the students agreed with Andy, 54$\%$  with Caroline and 22$\%$ with neither (correct response). Indeed this difficulty is quite common even amongst graduate students and graduate level instruction does not help students develop a better understanding of these concepts.

 \begin{table}
  \centering
 \caption[1]{\parbox[t]{9.5cm}{The answer distributions of 17 graduate students in the first semester and 15 graduate students in the second semester answering the same question about energy measurement.}}
  \begin{tabular}{|c|c|c|c|c|c|c|}
    \hline
 Answers    & \ \ \ A \ \ \ & \ \ \ B \ \ \  & C(correct) & \ \ \ D \ \ \  & \ \ \ E  \ \ \ & No Answer \\
    \hline
First Sem.  & \ \ \ 47$\%$ \ \ \	& \ \ \ 12$\%$ \ \ \ & 29$\%$ & \ \ \ 0$\%$ \ \ \	& \ \ \ 0$\%$ \ \ \ & 12$\%$ \\
    \hline
Second Sem. & \ \ \ 20$\%$ \ \ \	& \ \ \ 13$\%$ \ \ \ & 53$\%$ & \ \ \ 7$\%$ \ \ \	& \ \ \ 0$\%$ \ \ \ & 7$\%$ \\
    \hline
  \end{tabular}
\end{table}

\section*{\uppercase\expandafter{\romannumeral3} . SUMMARY AND CONCLUSION}

We find that students share common difficulties with concepts related to quantum measurement. In particular, many students were unclear about the difference between energy eigenstates and eigenstates of other physical observables and what happens to the state of the system after the measurement of an observable. Students also had difficulty in distinguishing between the measured value, the probability of measuring it and the expectation value. They often did not think of the expectation value of an observable as an ensemble average of a large number of measurements on identically prepared systems but rather thought of it as a mathematical procedure where an operator is sandwiched between the same bra and ket states (the state of the system) or the integral formulation for calculating the expectation value in the position representation. Students were also confused about whether the system is stuck in the state in which it collapsed right after the measurement or whether it goes back to the state before the measurement was performed. In general, students struggled with issues related to the time evolution of wave function after the measurement. Based on the investigation of students' difficulties, we developed the Quantum Interactive Learning Tutorial (QuILT) and concept tests to improve students' understanding of quantum measurement. These research-based learning tools will be discussed in the second of the two papers (Part 2).

\begin{acknowledgments}
This material is based upon work supported by the National Science
Foundation. 
\end{acknowledgments}


\begin{thebibliography}{00}
\bibitem{bib1}D. J. Griffiths, \emph{Introduction to Quantum Mechanics} (Prentice Hall, Upper Saddle River, NJ, 1995).
\bibitem{bib2}L. D. Carr and S. B. McKagan, Graduate quantum mechanics reform, Am. J. Phys. \textbf{77}, 208 (2009).
\bibitem{bib3} D. Styer, \emph{The strange world of quantum mechanics} (Cambridge Univ. Press, London, 2000).
\bibitem{add}G. Zhu and C. Singh, Surveying students? understanding

of quantum mechanics in one spatial dimension, Am.

J. Phys. 80, 252 (2012); S. Y. Lin and C. Singh,

Categorization of quantum mechanics problems by pro-
fessors and students, Eur. J. Phys. 31, 57 (2010); A. Mason

and C. Singh, Do advanced physics students learn from

their mistakes without explicit intervention?, Am. J. Phys.

78, 760 (2010).
\bibitem{bib4}P. Jolly, D. Zollman, S. Rebello, and A. Dimitrova, Visualizing potential energy diagrams, Am. J. Phys. \textbf{66}, 57 (1998); D. Zollman, S. Rebello, and K. Hogg, Quantum physics for everyone: Hands-on activities integrated with technology, Am. J. Phys. \textbf{70}, 252 (2002).
\bibitem{bib5}D. Styer, Common misconceptions regarding quantum mechanics, Am. J. Phys. \textbf{64}, 31 (1996).
\bibitem{bib6}I. D. Johnston, K. Crawford, and P.R. Fletcher, Student difficulties in learning quantum mechanics, Int. J. Sci. Educ. \textbf{20}, 427 (1998); G. Ireson, The quantum understanding of pre-university physics students, Phys. Educ. \textbf{35}, 15 (2000).
\bibitem{bib7}S. B. McKagan, K. K. Perkins, and C. E. Wieman, Deeper look at student learning of quantum mechanics: The case of tunneling, Phys. Rev. ST Phys. Educ. Res. \textbf{4}, 020103 (2008); S. B. McKagan, K. K. Perkins, and C. E. Wieman, Design and validation of the Quantum Mechanics Conceptual Survey, Phys. Rev. ST Phys. Educ. Res. \textbf{6}, 020121 (2010).
\bibitem{bib8}M. Wittmann, R. Steinberg, E. Redish, Investigating student understanding of quantum physics: spontaneous models of conductivity, Am. J. Phys. \textbf{70}, 218 (2002).
\bibitem{bib9} C. Singh, Student understanding of quantum mechanics, Am. J. Phys. \textbf{69}, 885 (2001).
\bibitem{bib10}C. Singh, Student understanding of quantum mechanics at the beginning of graduate instruction, Am. J. Phys., \textbf{76}, 277 (2008).
\bibitem{bib11}Research on Teaching and Learning of Quantum Mechanics, papers presented at the National Association for Research in Science Teaching, perg.phys.ksu.edu/papers/narst/ (1999); also see theme issue of Am. J. Phys. \textbf{70}, 2002 published in conjunction with Gordon Conference on physics research and education: quantum mechanics.
\bibitem{bib12}C. Singh, M. Belloni, and W. Christian, Improving student's understanding of quantum mechanics, Physics Today \textbf{8}, 43 (2006).
\bibitem{bib13}G. Zhu and C. Singh, Improving students' understanding of quantum mechanics via the Stern-Gerlach experiment, Am. J. Phys. \textbf{79}, 499 (2011); G. Zhu and C. Singh, Students' Understanding of Stern Gerlach Experiment, in \emph{Phys. Ed. Res. Conference}, edited by C. Henderson, M. Sabella, and C. Singh (Ann Arbor, MI, 2009), Vol. 1179, p. 309.
\bibitem{bib14}C. Singh, Interactive learning tutorials on quantum mechanics, Am. J. Phys. \textbf{76}, 400, (2008).
G. Zhu and C. Singh, following article, Improving stu-
dents understanding of quantum measurement. II.
Development of research-based learning tools, Phys.
Rev. ST Phys. Educ. Res. 8, 010118 (2012)
\bibitem{bib15}M. Chi, Thinking Aloud, in \emph{The Think Aloud Method: A Practical Guide to Modeling Cognitive Processes}, edited by M. W. Van Someren, Y. F. Barnard, and J. A. C. Sandberg (Academic, London, 1994).
\bibitem{bib16}C. Singh, Student difficulties with quantum mechanics formalism, in AIP Conf. Proc. \textbf{883}, 185 (2007).
\bibitem{bib17}M. Wittmann, J. Morgan, R. Feeley, Laboratory-tutorial activities for teaching probability, Phys. Rev. ST PER \textbf{2}, 020104 (2006).
\end{thebibliography}
\end{document}